\documentclass{iaus}
\usepackage{graphicx}

\title[Probing extrasolar planet atmosphere and exosphere] 
{The extrasolar planet atmosphere and exosphere:  Emission and transmission spectroscopy}

\author[Tinetti \& Beaulieu]   
{Giovanna  Tinetti$^{1,2}$
 \and Jean-Philippe  Beaulieu$^{2,3}$}

\affiliation{$^1$University College London, Gower street, London WC1E 6BT, UK
 \\ email: {\tt g.tinetti@ucl.ac.uk} \\[\affilskip]
 $^2$ HOLMES collaboration \\[\affilskip]
$^3$Institut d'Astrophysique de Paris, 98bis Boulevard Arago, 75014 PARIS, France.
\\email: {\tt beaulieu@iap.fr}}

\pubyear{2008}
\volume{253}  
\pagerange{231--238}
\jname{Transiting planets}
\editors{Pont F., Queloz D., Sasselov., Torres M. and Holman M. eds}
\begin{document}

\maketitle

\begin{abstract}

We have entered the phase of extrasolar planets characterization, probing their
atmospheres for molecules, constraining their horizontal and vertical temperature profiles and estimating the contribution of  clouds and hazes. We report here a short review of the current situation using ground based and space based observations, and present the transmission spectra of HD189733b in the spectral range 0.5-24 microns.

\end{abstract}

\firstsection 

\section{Introduction}
Extrasolar Giant Planets (EGPs) are now being discovered at an ever increasing pace
(Butler et al., 2007, Schneider, 2008). And as a result, planetary
scientists and astronomers are increasingly called upon to make the transition from {\it
discovery} to {\it characterization}, so that we can begin the long task of
understanding these planets in the same way that we understand those in our own Solar System.
For a growing sample of giant extrasolar planets orbiting very close to their parent star, we can already probe
their atmospheric constituents using transit techniques. A stellar occultation (called primary transit) occurs when the light from a star is partially blocked by an intervening body, such as a planet, from reaching an observer. With this method, we can indirectly
observe the thin atmospheric ring surrounding the optically
thick disk of the planet -the limb- while the planet is transiting in
front of its parent star. This method was traditionally  used to probe the
 atmospheres  of  planets in our Solar System and most recently, thanks to
the  {\it Hubble Space Telescope} and {\it Spitzer}, was successfully applied to a
growing sample of giant exoplanets orbiting very close to their parent
star - the so called "Hot-Jupiters"-. The idea was first theoretically
proposed by Seager and Sasselov in 2000, and confirmed experimentally by
Charbonneau {\it et al.} in 2002, when he and his team first detected the
presence of Sodium in the atmosphere of a hot-Jupiter.

In the secondary transit technique, we observe first the combined spectrum
of the star and the planet. Then, we take a second measurement of the star
alone when the planet disappears behind it: the difference between the two measurements
consists of the planet's contribution. This technique was pioneered by two different teams in 2005,
using the Spitzer Space Telescope to probe two Hot-Jupiters in the Infrared (Deming et al., 2005; Charbonneau et al., 2005). A similar measurement was performed in 2006 by Harrington and collaborators on a non-transiting planet by monitoring through time the combined star-planet
flux. The light-curve obtained in this way, allowed to understand some thermal properties of the planet Upsilon Andromedae b.

{\bf Primary transit. } The use of transmission spectroscopy to probe the outer layers of the transiting hot Jupiters
has been particularly successful in the UV and visible spectral ranges (Charbonneau et al., 2002;
Vidal-Madjar et al., 2003; Ballester, Sing \& Herbert 2007; Ben Jaffel 2007, 2008;
Knutson et al., 2007a, Pont et al., 2007, Redfield et al., 2008)
and only more recently it was attempted in
 the Near and Middle Infrared spectral window, producing novel and extremely interesting results (Richardson et al., 2006, Deming et al. 2007, Knutson et al., 2007b, 2008a, Beaulieu et al., 2008,
Tinetti et al., 2007, Swain et al., 2008a, Nutzman P., et al., 2008, Agol et al., Knutson et al.,
Harington et al. this volume).
Transmission spectra are sensitive to atomic and molecular abundances and less to temperature variation. Temperature
influences the transmission spectrum by way of its influence on the atmospheric scale height (Brown 2001) and the absorbtion coefficients.

{\bf Secondary transit.}
With this method we can probe the photons that are directly emitted (Charbonneau et al., 2005, Deming et al., 2005), or reflected by the
planet (Cameron et al. 1999, Leigh et al., 2003, Rowe et al., 2006). So far, the focus has been on the brightest stars with transiting extrasolar planets,
namely HD 209458b (Charbonneau et al., 2000), HD 189733b (Bouchy et al., 2005)
and GJ436b (Butler et al., 2004), HD 149026b (Sato et al., 2005),
TRES-1 (Alonso et al., 2004).

In the infrared spectral range,  with this technique we can not only detect the molecular species showing a noticeable
rotational/vibrational signature,
but also constrain the bulk temperature and the vertical thermal gradient  (Knuston et al., 2007b, 2008b; Burrows et al., 2007;
Charbonneau 2008, Barman 2008, Harrington et al. 2006, 2007; Swain et al., 2008b).
Compared to transmission spectroscopy, emission spectroscopy may scan different regions of
the atmosphere for molecular signatures and cloud/hazes contributions (Brown, 2001; Richardson et al., 2007).
Same considerations are valid in the UV-visible spectral range, except that the photons reflected by the planet do not bring any information about the planetary temperature and the thermal structure, but about the planetary albedo  (Rowe et al., 2006) and the presence of atomic/ionic/molecular species having electronic transitions.

Finally, especially if the planet is tidally locked,
 with primary and secondary transit techniques we can observe different phases of the planet along
its orbit. During the primary transit we can sound the terminator,
whereas during secondary we can above all observe  the day-side.

{\bf Light-curves. }
Monitoring the light-curve of the combined star-planet spectrum, can be a useful approach both for transiting (Knutson et al., 2007, 2008) and non-transiting planets
(Harrington et al., 2006). In the latter case the planetary radius can not be measured, but we can appreciate the temperature or albedo variations through time (depending if the observation is performed in the visible or infrared).
\newpage
\newpage
The problems that we can tackle with current telescopes are :
\begin{itemize}
\item Detection of the main molecular species in the hot transiting planets' atmosphere.
\item Constraint of the horizontal and vertical thermal gradients in the hot exoplanet atmospheres.
\item Presence of clouds or hazes in the atmospheres.
\end{itemize}
Today we can use two approaches to reach these objectives:
\begin{enumerate}
\item Broad band or low resolution spectroscopy from a space based observatory. This can be
 accomplished by SPITZER or HST.
\item High resolution spectroscopy from ground based observatories in the optical and NIR.
\end{enumerate}
The next steps with these indirect techniques will be:
\begin{itemize}
\item Detection of minor atmospheric species and constraint of their abundance.
\item More accurate spectral retrieval to map thermal and chemistry gradients in the atmospheres.
\item Cloud microphysics: understanding the composition, location and optical parameters of cloud/haze particles.
\item Cooler and smaller planets, possibly in the habitable zone.
\end{itemize}
Futher into the future the James Webb Space Telescope or the  JAXA/ESA SPICA mission concept (Nakagawa et al., 2003)   will be the next generation of space telescopes to be online. They will guarantee high spectral resolution from space and the characterization of fainter targets, allowing us to expand the variety of "characterizable" extrasolar planets.

\section{Temperature profiles}
With photometry, or low resolution spectroscopy in the Near and Mid Infrared we are today
able to put some constraints on the thermal horizontal and vertical profiles of the planetary
atmospheres. For instance, we are already in a position of appreciating the differences
between HD189733b (Knutson et al., 2007b) and Upsilon Andromedae (Harrington, et al., 2006):
at 8and 24 $\mu m$ HD189733b shows a well mixed temperature distribution between the day and the
night side, while we have the opposite for Upsilon Andromedae at 24 $\mu $m.
HD209458b shows clear signs of a thermal inversion at relatively low altitude
(Burrows et al., 2007), the situation is different for  HD189733b  (Barman 2008, Swain et al., 2008b).
However high resolution spectroscopy is needed to perform a more accurate spectral retrieval
and better constrain the dynamics of planetary atmosphere.

\begin{figure}
\includegraphics[width=10cm,angle=90]{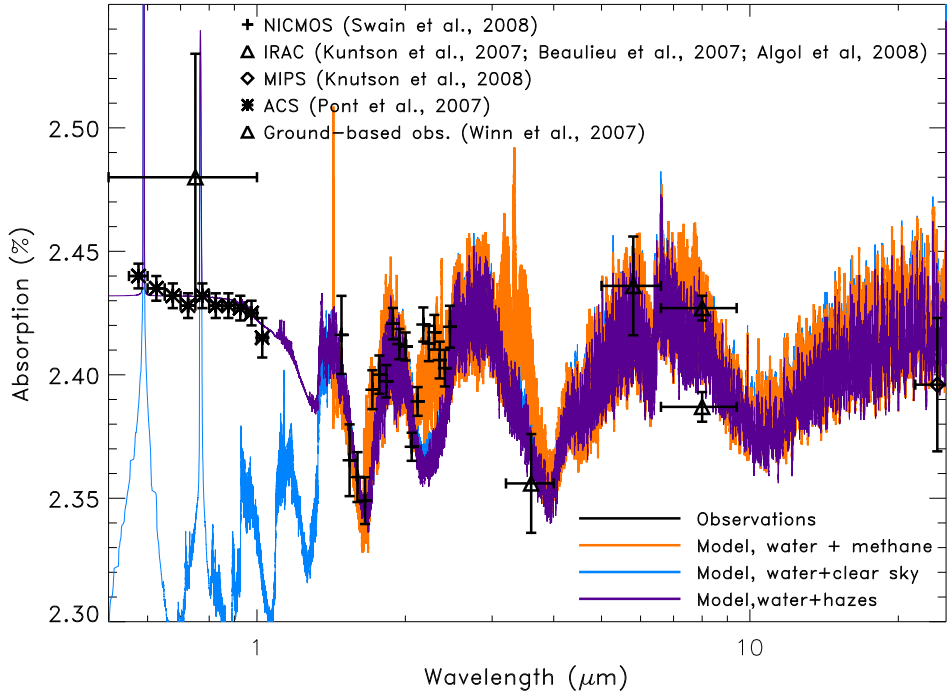}
\caption{Simulated middle Infrared spectra of the transiting Hot Jupiter HD189733b in the wavelength range 0.5-25 $\mu$m together
with HST ACS Measurements in the range $0.5-1 \mu$m (Pont et al., 2007), HST NICMOS at $1.5-2.5 \mu$m (Swain, Vasisht, Tinetti, 2008), and Spitzer IRAC-MIPS measurements (Beaulieu et al., 2008; Knutson et al., 2007, 2008; Agol et al., this volume). The overall transmission spectrum is shaped by the water absorption in
the infrared (Tinetti et al., 2007b) but methane is needed to explain the NIR (Swain, Vasisht, Tinetti, 2008). Notice that the different data collected by instruments over a wide wavelength range
are giving consistent results. At shorter wavelength, the increasing flatness of the spectrum could be explained by hazes. In our simulation we have used a distribution of haze particles with size smaller than 1 $\mu$m. To better constrain the optical properties of the
hazes, we should have additional data between 1 and 1.6 $\mu$m.  With the current data, we cannot make any conclusive hypothesis about the haze chemical composition.
}
\end{figure}

\section{Molecules in the atmosphere of hot-Jupiters}

\noindent {\bf Water.} In a star-planet system, a significant amount of water vapor can only exist
in planetary atmospheres at orbital distances small enough (less than $\sim 1 $AU for a solar
 like star).  This closeness requirement is well met by hot Jupiters.
According to photochemical models, H$_2$O should be among the most abundant species (after H$_2$) in the lower
atmosphere of giant planets orbiting close to their parent stars (Liang et
al., 2003; 2004).
Moreover, according to our calculations (Tinetti et al., 2007a), H$_2$O is the easiest of these species to detect in primary transit in
the IR.  A very accurate data list for water at hot Jupiter-like temperatures, has
 been calculated by Barber et al. (2006).

\noindent {\bf Carbon bearing molecules}, such as $CO$, $CH_4$, $CO_2$, $C_2H_2$ are expected to be abundant as well, depending on the C/O ratio and the efficiency of the
photochemistry in the upper atmosphere (Liang et al., 2004). $CO$, $CH_4$, $CO_2$ have already been detected (Barman, 2008, Charbonneau et al. 2008, Swain et al., 2008a, Swain et al., 2008b).
For less abundant species, or with spectral signatures which are harder to detect, we need to make the leap to high resolution spectroscopy.

Also, improved line lists at high temperatures are needed to better interpret the measurements.
From preliminary results and models, it is not excluded that the chemistry might vary substantially from the highly irradiated day-side to the non-illuminated night-side of these planets (Cooper and Showman, 2006; Swain et al., 2008a). 

\noindent {\bf Nitrogen or sulfur-bearing molecules}  are also likely to be present in the atmospheres of hot-Jupiters, but their weaker signatures may be difficult to be caught with a low resolving power (Sharp and Burrows, 2007).

\noindent {\bf H$_3^+$} -- the simple molecular ion formed by the photo-ionisation
of H$_2$ -- could be a crucial indicator of the escape processes
in the upper atmosphere (Yelle, 2004, 2006). Now that HST/STIS instrument is no
 longer operative to observe the Lyman alpha line in the UV (Vidal-Madjar et al. 2003, Ben-Jaffel, 2007), H$_3^+$ is the
only molecular ion able to monitor the escape processes on hot Jupiters. So H$_3^+$
is a crucial detection target; even if detection is unsuccessful, such measurement
provides at least an improved upper limit on its abundance (Shkolnik et al., 2006).
Calculations of the H$_3^+$ abundance on a hot-Jupiter (Miller et al., 2000; Yelle, 2004)  show that the contrast between H$_3^+$  emission and stellar brightness places it just on the current limit of detectability with a large ground-based telescope. We stress that even if it is a very challenging observation, it would be the best diagnostics to understand
the properties of the upper atmospheres of Hot-Jupiters (Koskinen et al., 2007) whereas observations of the Lyman alpha line could be partially or totally contaminated by energetic neutral atoms from charge exchange between stellar wind protons and neutral hydrogen from the planetary exospheres (Holmstrom et al., 2008) or inadequately analyzed and understood as stressed by Ben-Jaffel (2008).

\noindent {\bf Clouds and hazes}. At the spectral resolution we can obtain today from space, the best we can do is to assess their presence, as they are supposed to flatten the spectral signatures or modify the spectral shape. In the case of transmission spectroscopy, they  cause the atmosphere to be opaque at higher altitudes (Brown, 2001). The HST observations from Pont et al., (2008) show an almost featureless transmission spectrum in the range $0.5-1 \mu $m which may suggest the presence
of hazes.

\begin{figure}
\includegraphics[width=10cm,angle=90]{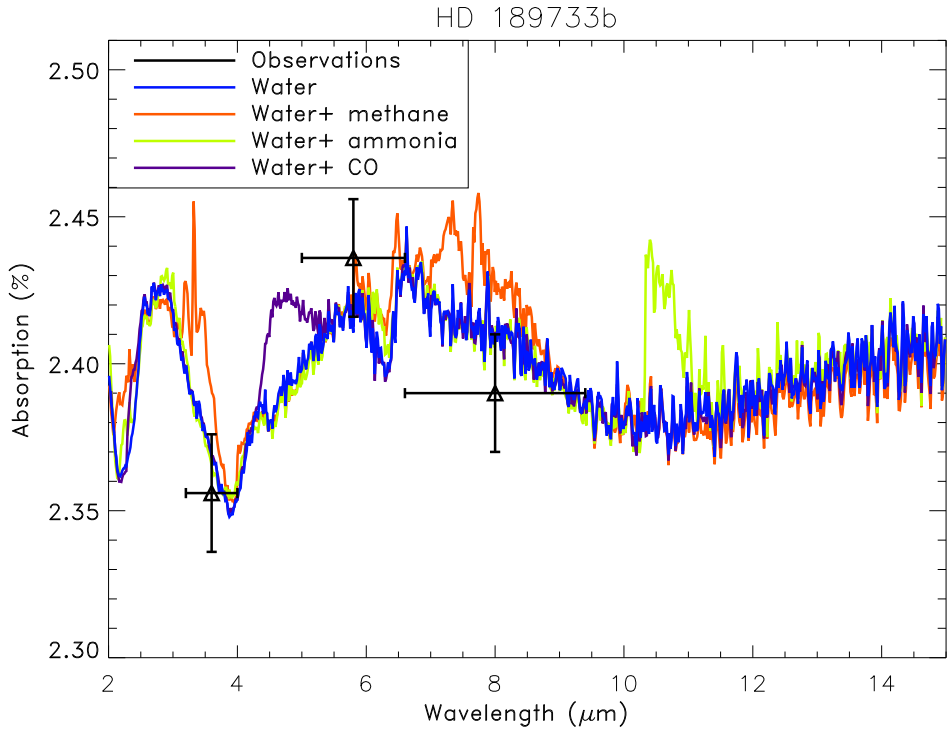}
\caption{Simulated middle Infrared spectra of the transiting Hot Jupiter HD189733b in the wavelength range 2-15 $\mu$m together
with IRAC measurements (Beaulieu et al., 2008, Knutson et al., 2007, Tinetti et al., 2007b).
Water is giving the main pattern of the spectra. Notice that to distinguish the different
additional molecules, photometry is not enough, highly resolved spectra are needed.
}
\end{figure}

As an illustration we present the transmission spectrum of HD189733b in the wavelength range
$0.5-25 \mu $m in Figure 1. The overall transmission spectrum is shaped by the water absorption in
the infrared. H2-H2, methane and alkali metals absorptions are included, as well as a crude simulation of hazes opacity. Notice that the different data collected by instruments over a wide wavelength range
are giving consistent results. Most probably additional molecules are present, but we are unable to appreciate their presence at this spectral resolution. For instance,
in Figure 3., we show the additional contribution of a variety of plausible molecules as a function of wavelength:  the contributions of Methane at 3.2 $\mu$m, CO at 4.5
$\mu$m and ammonia at 11 $\mu$m are quite noticeable. Although water can be detected
 with broad band photometry, it is clear that spectral resolution is needed in order to get the probe for the different species.

\section{Towards smaller mass planets}

For both primary and secondary transit methods,  smaller size/colder planets will increase
the challenge. Transmission spectroscopy can benefit from very extended atmospheres : this scenario can occur if the main atmospheric component has a light molecular weight and a high temperature. The lighter, the hotter and the smaller the core, the easier is the observation in transmission spectroscopy (ie the more detectable are the spectral features).
In the case of secondary transits, the parameters playing the major role are the
size of the planet compared to its parent star and the planetary temperature for observations in the IR or the albedo in the visible.

With current telescopes we can already approach the case of hot Neptunes transiting later type stars, e.g. Gliese 436b (Deming et al., 2007). Figure 3 shows a simulated transmission spectrum of Gliese 436b. Note that even if the transit depth is smaller than
a hot Jupiter one
(0.7 \%), molecules such as methane could leave signatures of similar order of magnitude
($\sim$0.05 \%).

\begin{figure}
\includegraphics[width=10cm,angle=90]{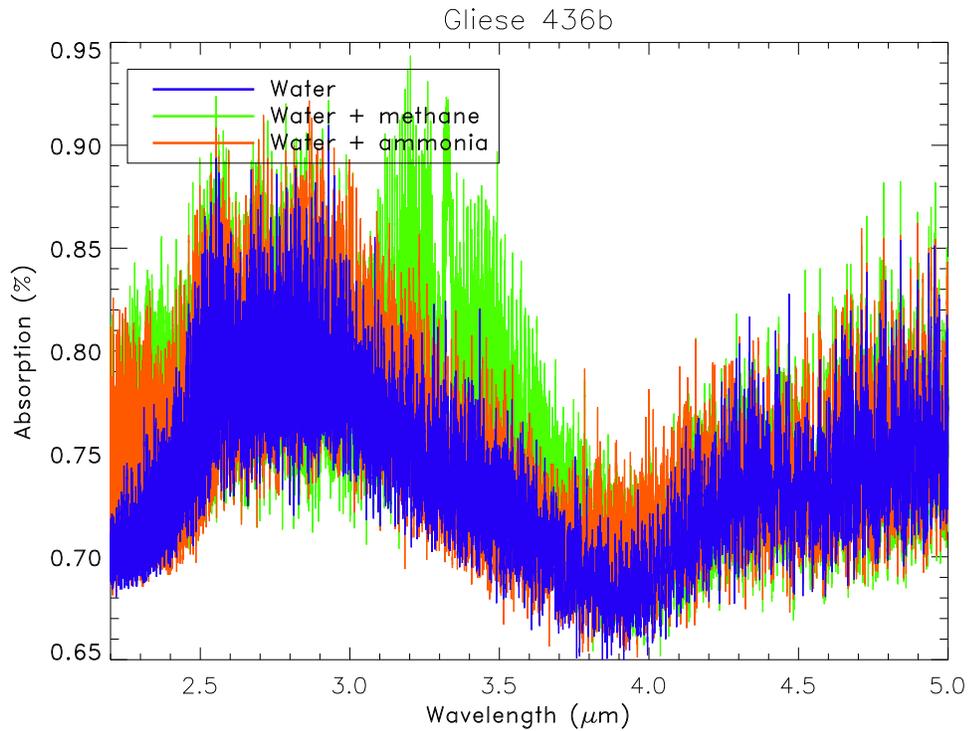}
\caption{Middle Infrared spectra of the transiting
Hot Neptune GJ436b in the wavelength range 2.5-4.5 $\mu$m. The blue curve is due to water alone, the red curve to water and methane. The Strongest signature of Methane
in the MIR is at 3.3-3.4 $\mu$m. Water mixing ratio is supposed to be $5~10^{-4}$, methane mixing ratio $10^{-5}$. If methane mixing ratio is decreasing or increasing, the red plot will accordingly change.
}
\end{figure}

For Earth-size planets and/or colder atmospheres, we need to wait for JWST (Cornia and Tinetti, 2007; Cavarroc et al., 2006, 2008).

\section{Conclusion}

Probing the exoplanet atmospheres with transiting techniques has a bright and exciting future, both from space and from the ground.
 There are two main approaches: primary and secondary transit methods. In this proceeding we have reviewed how they can be complementary, and what are their inherent limitations. Taken together, and over a broad spectral range, these methods allow us to reach for exoplanet atmospheres similar level of knowledge that scientists had of the planets in the solar system at the time of Voyager 1.

\noindent{\bf Acknowledgements}\\
  It is a great pleasure to thank our collaborators for many exciting and fruitful discussions, in particular Mark Swain, Gautam Vasisht, Ignasi Ribas, Sean Carey, Jeroen Bouwman, Danie Liang, Yuk Yung, Eric Agol, Jonathan Tennyson, Alan Aylward, Bob Barber, Steve Miller, Virginie Batista, Pieter Deroo, David Kipping and Tommi Koskinen. J.P.B. and G.T. acknowledge the financial support of the ANR HOLMES.


\end{document}